\documentclass[aps,prl,twocolumn,showpacs]{revtex4-1}  
\usepackage[version=3]{mhchem}	
\usepackage{graphicx}  
\usepackage{nicefrac}  
\usepackage{siunitx} 	
\usepackage{amsmath,amssymb,amsfonts}   
\usepackage{hyperref}
\usepackage{import}
\usepackage{bbold}
\usepackage{wasysym}
\usepackage{dcolumn}
\usepackage{afterpage}
\usepackage{pdfpages}

\hypersetup{colorlinks=true,urlcolor=blue,citecolor=blue,linkcolor=blue,breaklinks}
\renewcommand{\vec}[1]{\mathbf{#1}}

\newcommand{\ie}{\emph{i.e.}}
\newcommand{\eg}{\emph{e.g.}}
\newcommand{\CR}{Cr\ensuremath{_2}O\ensuremath{_3}}

\newcommand{\alp}{\ensuremath{\vec{\alpha}}}

\newcommand{\alatt}{\ensuremath{\boldsymbol{\alpha}^\text{latt}}}
\newcommand{\PGC}{\ensuremath{\bar{3}'m'}}
\newcommand{\PGO}{\ensuremath{2'/m}}
\newcommand{\PGT}{\ensuremath{2/m'}}
\newcommand{\atu}{\ensuremath{A_{2u}}}
\newcommand{\eu}{\ensuremath{E_{u}}}
\newcolumntype{d}[1]{D{.}{.}{#1}}

\begin{document}

\title{\emph{Ab Initio} Cycloidal and Chiral Magnetoelectric Responses in \CR}
\author{Natalie Tillack}
\email{natalie.tillack@materials.ox.ac.uk}
\affiliation{Department of Materials, University of Oxford, Oxford, OX1 3PH, United Kingdom}
\author{Paolo G. Radaelli}
\affiliation{Department of Physics, University of Oxford, Oxford, OX1 3PU, United Kingdom}
\author{Jonathan R. Yates}
\affiliation{Department of Materials, University of Oxford, Oxford, OX1 3PH, United Kingdom}

\date{\today}

\begin{abstract}
We present a thorough density functional theory study of the magneto-electric (ME) effect in \CR. The spin-lattice ME tensor \alp\ was determined in the low-field and spin flop (SF) phases, using the method of dynamical magnetic charges, and found to be the sum of three distinct components. Two of them, a large relativistic ``cycloidal'' term and a small longitudinal term, are independent on the spin orientation. The third, only active in the SF phases is also of relativistic origin and arises from magnetic-field-induced chirality, leading to a non-toroidal ME response.
\end{abstract}

\maketitle

\begin{figure}[htpb]
  \centering
  \scriptsize{
\def\svgwidth{0.8\columnwidth}
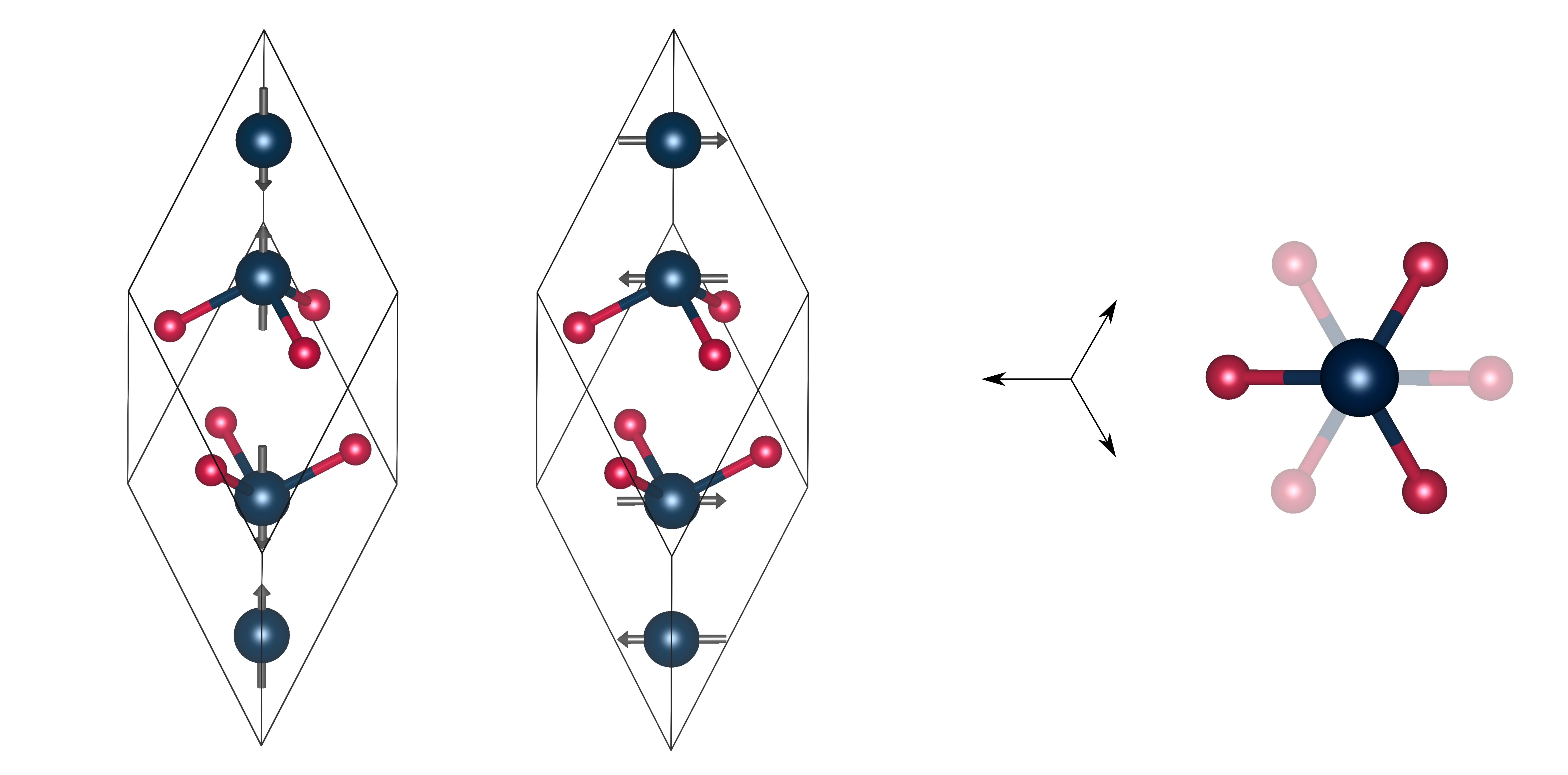}
\caption{\label{fig:structure}Structures of \CR. Left panel: Rhombohedral primitive cell of \CR\ with arrows indicating the AFM coupled \ce{Cr} magnetic moments along $z$. The magnetic point group is $\bar{3}'m'$. Two $\SI{180}{\degree}$ domains are possible, linked via time reversal. Middle panel: The spin flop state ($2'/m$ or $2/'m$), with spins aligned in the basal plane, thus breaking the 3-fold symmetry. Right panel: View along $z$ (the half-transparent \ce{O} atoms belong to the ``lower'' structural unit). Figures plotted with VESTA \cite{Momma2011}.}
\end{figure}
\begin{figure}[htpb]
  \centering
  \scriptsize{
\def\svgwidth{0.8\columnwidth}
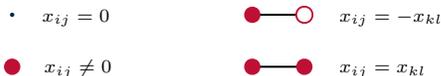}
\caption{\label{fig:tensor_forms}Tensor forms of the BEC $Z^e$ and magnetic charges $Z^m$ (both based on the atom's site symmetry) and of the overall ME coupling tensor~\alp~(based on the point group of the magnetic crystal class).}
\end{figure}

The search for magneto-electric (ME) materials, in which the electrical polarisation $P$ (the magnetisation $M$) responds to the application of an external magnetic field $H$ (electric field $\epsilon$), has received a lot of attention in recent years \cite{Schmid1994,Fiebig2005,Cheong2007,Scott2012}, particularly in the context of `modern' multiferroic materials with a spontaneous polarisation \cite{Eerenstein2006}. The linear magneto-electric effect, whereby $P$ is linearly proportional to $H$, is also of current technological interest for magnetic storage devices, replacements of SQUIDs, and the ME switching of exchange bias \cite{Fiebig2009,Borisov2005,Borisov2007}. In the 1950s, Landau and Lifshitz were the first to demonstrate that the ME effect only occurs in magnetic (\ie, time-reversal odd) materials \cite{Landau}. \CR, often considered the prototypical ME, crystallises in the trigonal corundum structure and, below the N\`eel temperature of $T_N=\SI{307}{\kelvin}$, orders as a collinear antiferromagnet (AFM) with spins along the rhombohedral $[111]$ direction (Fig. \ref{fig:structure}). \CR\ was predicted to be magneto-electric based on symmetry considerations \cite{Dzyaloshinsky1958,Dzyaloshinskii1960} --- a prediction that was later verified experimentally \cite{Astrov1960,Astrov1961,Folen1961,Rado1962,Rado1962a}. Unlike most other MEs, \CR~ is ME above room temperature, making it technologically relevant in spite of the small ME response \cite{Street2014}. \CR~ is also ideal for studying the fundamental ME mechanisms, since it is not multiferroic, and -- because of its magnetic point group -- exhibits neither higher-order ME coupling nor piezomagnetism. Nevertheless, there is still a surprising amount of uncertainty surrounding the ME effect in \CR, and in particular its behaviour throughout the T-H phase diagram; in turns this hampers the systematic search for materials with a stronger ME response. \\
The linear ME coupling can be described by an axial tensor of rank two: 

\begin{equation}
\alpha_{ij}  = \left( \frac{\partial P_i}{\partial H_j} \right) = \mu_0  \left( \frac{\partial M_j}{\partial \epsilon_i} \right) 
\end{equation} 

\noindent with $\vec{P}$ ($\vec{M}$) being the induced polarisation (magnetisation), $\vec{H}$ ($\boldsymbol{\epsilon}$) the external magnetic (electric) field, and $\mu_0$ the magnetic permittivity. The
components of P and H are conventionally expressed in a Cartesian coordinate system, with $z$ along the rhombohedral $[111]$ direction, $x$ along one of the 2-fold axes, and $y$ completing the right-handed set. The form of the linear ME tensor $\alpha_{ij}$ can be predicted entirely by symmetry once the AFM point group is known \cite{Wiegelmann1994, Borovnik-Romanov2006}. In low applied H (LF phase), the \CR\ spins are aligned along $z$ due to magnetic anisotropy \cite{Allen1973,Gibbs2001} (magnetic point group $\bar{3}'m'$), making $\alpha_{ij}$ diagonal and $\alpha_{11}$ and $\alpha_{22}$ being equal (Fig. \ref{fig:tensor_forms}). $\alpha_{33}$ is very small in the ground state, but becomes the dominant element at room temperature \cite{Kita1979,Scaramucci2012}. Under strong applied fields along $z$, \CR\ undergoes a first-order phase transition into the so-called \emph{spin flop} (SF) phase, with spins ordered in the same $G$-type pattern, but directed in the basal plane \cite{Ohtani1984} (middle panel in Fig.~\ref{fig:structure}). The possible magnetic point groups of the SF phase, \PGO, \PGT, or $\bar{1}'$ for spins \emph{parallel} or \emph{perpendicular} to $x$ or in a generic direction, respectively, also allow for the ME effect, but with a different, off-diagonal form of the ME tensor (see Fig. \ref{fig:tensor_forms}), which is indeed observed experimentally \cite{Popov1999}. The ME effect in the SF phase has often being associated with the appearance of a \emph{toroidal moment} $\boldsymbol{T}=\sum_{i} \bf{r}_i \times \bf{S}_i$ \cite{Popov1999, Spaldin2008}. However due to the presence of multiple domains in the SF phase (6 domains are predicted, due the 3-fold symmetry breaking), it is unclear whether the ME tensor is purely toroidal (antisymmetric). 

In the Letter, we probe the ME effect in the LF and the in two high-symmetry SF phases (\PGO\ and \PGT) in the ground state of \CR\ by a set of highly-controlled first-principle calculations --- an approach that yields results that are fully consistent with experiments but avoids the domain problem. We demonstrate that the `large' components of the ME tensor in the SF phase are not intrinsically toroidal. Rather surprisingly, these components are numerically identical to the $\alpha_{11}$ and $\alpha_{22}$ tensor elements in the LF phase, clearly indicating a common origin. We further show that the signs and identical magnitudes of these `large' ME components can be predicted from the cycloidal spin-current mechanism, which is well known in multiferroics. Finally, we show that the `small' ME components in the SF phase arise from two separate mechanisms: a longitudinal response (\eg, $\alpha_{33}$ in the LF phase), which in the ground state is associated with a small longitudinal susceptibility of relativistic origin, whereas it becomes the dominant response at room temperature \cite{Mostovoy2010}, and a novel chiral ME coupling.

The ME response of \CR\ in the LF phase has been the subject of a number of first-principles studies \cite{Iniguez2008,Mostovoy2010,Coh2011,Bousquet2011,Bousquet2011a,Malashevich2012,Ye2014}. The spin-lattice response \alatt\ has been shown to be dominant~\cite{Malashevich2012}, and we therefore focus on this contribution. We take the approach of Ref.~\cite{Ye2014} and expand the macroscopic response into microscopic quantities as follows:

\begin{equation}
 \alpha_{kl}^\text{latt} = \frac{\partial P_k}{\partial H_l} = \left( \frac{\partial P_k}{\partial u_i}\right) \left( \frac{\partial u_i \partial u_j}{\partial E} \right) \left( \frac{\partial F_j}{\partial H_l}\right)
\label{eq:alatt_method}
\end{equation}

\noindent with the indices $k,l=1,2,3$ and the composite indices (accounting for three directional dimensions and the number of atoms in the unit cell $N$) $i,j=1,\ldots,3N$. Eq.~\ref{eq:alatt_method} shows a trilinear relation involving the Born effective charges (BEC) $Z_{k i}^e=\frac{\sigma}{e} \frac{dP_k}{du_i}= - e \frac{dF_{i}}{d\epsilon_k}$, the inverse of the force-constant (FC) matrix $K_{ij}^{-1}= \frac{\partial u_i}{\partial F_j} = \frac{\partial u_i \partial u_j}{\partial E} $, and the dynamical magnetic charges (MC). The latter can be understood as the magnetic analogue of the BEC and are defined as the derivative of the Hellman-Feynman forces with respect to a magnetic field, $Z_{jl}^m=\frac{\partial F_j}{\partial B_l}=\mu_0^{-1}\frac{\partial F_j}{\partial H_l}$, using $\vec{B}=\mu \vec{H} \approx \mu_0 \vec{H}$ and the permeability (vacuum permeability) $\mu$ ($\mu_0$) being $\mu \approx \mu_0$ in AFMs. 

We calculate the three contributions to Eq.~\ref{eq:alatt_method} using Density Functional Theory within the local density approximation. We find that spin-orbit coupling (SOC) has a very small effect on the BEC and FC matrix (less than  \SI{1}{\permil}) and we therefore compute these quantities without SOC. This means that our BEC and FC matrix are the same in all three magnetic phases. In contrast, the MC are a SOC induced effect and we compute them using a non-collinear magnetism formalism employing SOC. A Zeeman magnetic field is applied according to Ref. \cite{Bousquet2011a} and the change in the ionic forces calculated. It is therefore the changes in the MC which determine the differing form of \alatt\ in the three magnetic phases. Full details of the DFT calculations, including an analysis of the influence of the choice of exchange-correlation functional are provided in the Supplementary Materials.


\begingroup
\squeezetable
\begin{table*}[htp]
\caption{\label{tab:cro_results_collected}Wyckoff positions, magnetic charges $Z^m$ [\SI{e-2}{\micro_B\per\angstrom}], and the overall ME coupling tensor \alp~[\si{\pico\second\per\meter}] for the LF and to SF phases of \CR. The \ce{O1} and \ce{O2} positions in the \PGC~phase are related by a 3-fold rotation and the $Z^m$ given in the following table for comparison with the SF cases.} 
\begin{ruledtabular}
\begin{tabular}{llllcccc}
Phase & $w(\ce{Cr})$ & $w(\ce{O1})$ & $w(\ce{O2})$ & $Z^m(\ce{Cr})$ & $Z^m(\ce{O1})$ & $Z^m(\ce{O2})$ & \alp \\
\PGC & $4c$ & $6e$ & $6e$ & \rule{0pt}{6ex}  $\left( {\begin{array}{d{3.1} d{2.1} d{2.2}} 2.4 & 5.6 & 0.0 \\  -5.6 & 2.4 & 0.0 \\  0.0 & 0.0 & 0.0 \\ \end{array} } \right)$ & %
$\left( {\begin{array}{d{2.1} d{2.1} d{2.2}} -2.8 & 0.0 & 0.0 \\  0.0 & -0.5 &-0.0 \\  0.0 & -2.7 & -0.0 \\ \end{array} } \right)$ & %
$\left( {\begin{array}{d{2.1} d{2.1} d{2.2}} -1.1 & 1.0 & 0.0 \\  1.0 & -2.2 & 0.0 \\  2.3 & 1.4 & 0.0 \\ \end{array} } \right)$ & %
$\left( {\begin{array}{d{3.3} d{2.3} d{2.4}} 0.310 & 0.000 & -0.001 \\ 0.000 & 0.310 & 0.000 \\ 0.000 & 0.000 & 0.005 \\ \end{array} } \right)$ \\
\PGO &  $8f$ & $4e$ & $8f$ & \rule{0pt}{6ex}  $\left( {\begin{array}{d{3.1} d{2.1} d{2.2}} 0.0 & -0.1 & -2.3 \\ 0.0 & 0.0 & 5.6 \\ 0.1 & 18.9 & 0.0 \\    \end{array} } \right)$ & %
$\left( {\begin{array}{d{2.1} d{2.1} d{2.2}} 0.0 & -1.5 & 3.0 \\ -0.3 & 0.0 & 0.0 \\ -0.1 & 0.0 & 0.0  \\ \end{array} } \right)$ & %
$\left( {\begin{array}{d{2.1} d{2.1} d{2.2}} 0.3 & 0.8 & 0.8 \\ 0.2 & -1.4 & -1.2 \\ -0.1 & -0.1 & -2.2 \end{array} } \right)$ & %
$  \left(  {\begin{array}{d{3.3} d{2.3} d{2.4}} -0.001 & -0.012 & -0.309 \\ 0.001 & -0.001 & 0.000 \\ 0.019 & 0.000 & 0.000 \end{array} }  \right) $ \\
\PGT &  $8f$ & $4e$ & $8f$ & \rule{0pt}{6ex}  $\left( {\begin{array}{d{3.1} d{2.1} d{2.2}} -0.1 & 0.0 & -5.6 \\ 0.0 & 0.0 & -2.3 \\ -18.8 & 0.2 & 0.0 \end{array} } \right)$ & %
$\left( {\begin{array}{d{2.1} d{2.1} d{2.2}} 1.7 & 0.0 & 0.0 \\ 0.0 & -0.3 & 0.1 \\ 0.0 & -0.1 & 2.5 \end{array} } \right)$ & %
$\left( {\begin{array}{d{2.1} d{2.1} d{2.2}} -0.7 & 0.3 & -1.2 \\ 1.3 & 0.2 & 2.2 \\ -0.1 & -0.1 & -1.3  \end{array} } \right)$ & %
 $\left( {\begin{array}{d{3.3} d{2.3} d{2.4}} -0.012 & 0.000 & 0.000 \\ 0.000 & -0.002 & -0.309 \\ 0.000 & 0.019 & 0.001  \end{array} } \right)$ \\
\end{tabular}
\end{ruledtabular}
\end{table*}
\endgroup

Our results for the LF \PGC~phase were benchmarked against Ref.~\cite{Ye2014} leading to almost identical results. A full comparison to literature values is given in the Supplementary Material. Table \ref{tab:cro_results_collected} shows our results for the magnetic charges and for the ME tensor for the LF \PGC~phase with spins parallel to $z$, and the two SF phases, \PGO\ and \PGT, with spins \emph{parallel} and \emph{perpendicular} to $x$, which is also the direction of the surviving 2-fold axis. The tensor forms for both quantities are in good agreement with our group theoretical analysis from Fig. \ref{fig:tensor_forms}. To within $\SI{\pm 0.001}{\pico\second\per\meter}$, which we take to be our computational uncertainty, the ME coupling tensors, are as predicted (bottom row of Fig.~\ref{fig:tensor_forms} and right column of table~\ref{tab:cro_results_collected}). 

Because in the \PGC\ phase all improper rotations are coupled to time-reversal symmetry, the MC are of the same form as the BEC. Therefore, those phonon modes that couple to the electric field, \ie, the infrared (IR) active modes, are also the ones that couple to the magnetic field. In the $R\bar{3}c$ space group, the IR active modes are the doubly degenerate \eu\ modes, which are active in the $xy$ plane, and the singly degenerate \atu\ modes, active along $z$. Not including the acoustic modes, the $\Gamma$-centred IR active modes are therefore

\begin{equation}
\Gamma^\text{IR} = 4 \eu + 2 \atu. 
\end{equation}

From a mode decomposition of the BEC and the MC, we observe changes in the magneto-active response when $x$ and $y$ are no longer equivalent. The degeneracy of the IR active $E_u$ modes is removed, and other modes become magneto-active in $x$, $y$, or $z$. We also find that the exceptionally large component in the $Z^m(\ce{Cr})$ (the $32$ and $31$ component in the~\PGO\ and~\PGT\ phase, respectively) maps onto magneto-active modes that are mutually exclusive to the IR active ones. That value has therefore no effect on the coupling tensor. Even though the magnetic charge tensors are quite dissimilar, this leads to the rather surprising similarity of the coupling tensors. The full mode decomposition of the BEC and the MC for the LF phase is given in the Supplementary Material. 

On this basis, we make the following important observations. Firstly, the ME tensor \alp\ is \emph{not} what one would expect from a toroidal moment. There is clearly a toroidal (antisymmetric) component, but this is identical in magnitude to the traceless symmetric component. This should not be particularly surprising, since the toroidal mechanism $\bf{P}=\boldsymbol{T} \times \bf{H}$ does not capture the large difference between the longitudinal and transverse susceptibilities. Secondly, the magnitudes of the `large' elements of the ME tensor is the same (within error) in the two SF phases \emph{and} in the LF phase. In fact, all these large ME responses can be approximated by the following compact expression:

\begin{equation}
\label{eq: cycloidal_component}
\alatt \approx \SI{0.31}{\pico\second\per\meter}  \left( \begin{array}{ccc} \hat{m}_z&0&-\hat{m}_x\\0&\hat{m}_z&-\hat{m}_y\\0&0&0 \end{array}\right) 
\end{equation}

\noindent where $\hat{m}_x$, $\hat{m}_y$ and $\hat{m}_z$ are the components of a unit vector parallel to the spin on the \ce{Cr}($4c$) atom at Wyckoff position 0.1590. A clue as to the origin of this tensor form is the fact the all these components correspond to a \emph{transverse} spin response in the plane containing both the spins and $z$, so that the rotated spins under the action of the magnetic field can be thought as forming a segment of a cycloid. Here below, we show that the tensor form in Eq. \ref{eq: cycloidal_component} is exactly the one predicted from the spin current \cite{Katsura2005} or inverse Dzyaloshinsky-Moriya (DM) \cite{Mostovoy2006} mechanisms, which are well known in the context of multiferroics. We write the transverse response of the magnetisation $\vec{m}_i$ on \ce{Cr} site $i$ as

 \begin{eqnarray}
\Delta \vec{m}_i&=&\chi^T\frac{\vec{m}_i\times \left( \vec{H} \times \vec{m}_i\right)}{m^2}\nonumber\\
&=&\chi^T\left( \vec{H}-\frac{\vec{m}_i\left(\vec{H} \cdot \vec{m}_i\right)}{m^2}\right)
 \end{eqnarray}

\noindent where $\chi^T$ is the transverse susceptibility. The inverse DM polarisation is:

\begin{eqnarray} \label{eq: magel_products}
\vec{P}&=&\mu \vec{r}_{12} \times \left( \vec{m}_1 \times \vec{m}_2\right)=2\mu \chi^T\vec{r}_{12} \times \left(\vec{m} \times \vec{H}\right)\nonumber\\
&=&2\mu \chi^T\left(\vec{m}\left(\vec{H} \cdot \vec{r}_{12}\right)-\vec{H}\left(\vec{m} \cdot \vec{r}_{12} \right)\right) 
\end{eqnarray}

\noindent where $\mu$ is a coupling constant and, in the case of \CR, $\vec{r}_{12} \parallel c$. With this, the ME tensor takes the form:

\begin{eqnarray}
\label{eq:  P_cycloidal}
\boldsymbol{\alp} &=& 2\mu\chi^T \left(\vec{m} \otimes \vec{r}_{12} - \vec{m} \cdot \vec{r}_{12} \mathbb{1}\right)\nonumber\\
&=&-2\mu\chi^T  \left( \begin{array}{ccc} m_z&0&-m_x\\0&m_z&-m_y\\0&0&0 \end{array}\right)
\end{eqnarray}

\noindent where $\otimes$ is the outer (tensor) product and $\mathbb{1}$ is the unit tensor. Eq. \ref{eq:  P_cycloidal} and Eq. \ref{eq: cycloidal_component} have exactly the same form, including the non-trivial sign of the tensor elements.

In all phases, the ME tensor αlatt has a small longitudinal term (i.e., with the field H parallel to the spins) of magnitude \SI{0.005}{\pico\second\per\meter} (\SI{0.019}{\pico\second\per\meter}) for the LF (SF) phase(s), which generates $P\parallel z$ in all cases.  In the ground state, this is associated with a small longitudinal susceptibility of relativistic origin, while at finite temperatures, the symmetric Heisenberg exchange makes this term become the dominant contribution to \alp\ in the LF phase, as shown in Ref. ~\cite{Mostovoy2010}. 

Considerably more interesting is the additional \emph{in-plane transverse} term of magnitude \SI{-0.012}{\pico\second\per\meter}, which is only present in the SF phases. Here, the field $H$ lies in the $xy$ plane perpendicular to the spins and generates $P \parallel x$, i.e., to the surviving 2-fold axis of the monoclinic structure. Phenomenologically, this term is associated with a \emph{fifth order} invariant of the form $(m_x^2-m_y^2) A_x-2m_xm_yA_y $ where $\vec{A}=(m_yH_x-m_xH_y) \vec{P}$ transforms as an axial (parity-even) vector. In the remainder, we show that this term is due to the breaking of axial symmetry upon SF magnetic ordering, coupled with the breaking of chiral symmetry upon application of a magnetic field in the in-plane transverse direction. 

As we already mentioned, SF magnetic ordering breaks the 3-fold symmetry. Consequently, the \emph{crystallographic} symmetry is also lowered, due to coupling of the staggered magnetisation with a structural order parameter, which has the transformation properties of an axial vector $\vec{A}$. Minimising the Landau free energy with respect to  $\vec{A}$ in the usual way, one obtains:

\begin{eqnarray}
A_x&=&\lambda (m_x^2-m_y^2)\nonumber\\
A_y&=&-\lambda (m_x m_y)\
\end{eqnarray}

\noindent where $\lambda$ is a (small) magneto-elastic coupling constant. It is noteworthy that $A_y=0$ in both of the SF phases we considered, so $\vec{A}$ is directed along $x$. Upon application of a magnetic field in the in-plane transverse direction, the rotated spins can be thought as forming a segment of a helix, which has the distinct chirality $\vec{r}_{12} \cdot \left( \vec{m}_1 \times \vec{m}_2\right)$. In analogy to the ferroaxial multiferroic mechanism \cite{Johnson2011,Johnson2012a}, we can therefore write the following phenomenological polarisation:

\begin{eqnarray}
P_x&=&\mu\lambda (m_x^2-m_y^2)\,  \vec{r}_{12} \cdot \left( \vec{m}_1 \times \vec{m}_2\right)\nonumber\\
&=&2\mu \chi^T\lambda (m_x^2-m_y^2)\, \vec{r}_{12} \cdot \left(\vec{m} \times \vec{H}\right)\nonumber\\
&=&2\mu\chi^T\lambda (m_x^2-m_y^2) \, \left(\vec{r}_{12} \times \vec{m} \right) \cdot \vec{H}\nonumber\\
P_y&=&0
\end{eqnarray}

yielding

\begin{equation}
\boldsymbol{\alp}=2\mu \chi^T\lambda (m_x^2-m_y^2)\vec{\hat{x}} \otimes \left(\vec{r}_{12} \times \vec{m}\right)
\end{equation}

where $\vec{\hat{x}}$ is a unit vector along $x$. Since $\left(\vec{r}_{12} \times \vec{m}\right) = (-m_y, m_x, 0)$, the ME tensor has the desired form

\begin{equation}
\label{eq:  P_chiral}
\boldsymbol{\alp}=-2\mu \chi^T\lambda m^2 \left( \begin{array}{ccc} m_y & m_x &0\\0&0&0\\0&0&0 \end{array}\right)
\end{equation}

\noindent which corresponds to the first-principle result. By comparing Eq. \ref{eq:  P_chiral} with Eq. \ref{eq:  P_cycloidal}, one can understand why the former, containing the small parameter $\lambda$, is considerably smaller than the latter.

In summary, we have computed the magneto-electric tensor of \CR\ in both low-field and high-field (spin flop) phases by means of highly-controlled first-principle calculations. We find that the ME tensor is not primarily toroidal, as previously speculated. Instead, its approximate form can be well predicted phenomenologically using the spin-current model, which is well known in multiferroics. There are two additional small components: the first is a longitudinal component known from previous studies of the low-field phase, while the second arises from a novel chiral mechanism, which is akin to the ferroaxial mechanisms in multiferroics. 

\begin{acknowledgments}
The work done at the University of Oxford was funded by EPSRC grants, number EP/J003557/1, entitled ``New Concepts in Multiferroics and Magnetoelectrics" and number EP/M020517/1, entitled ``Oxford Quantum Materials Platform Grant'', the Scatcherd European Scholarship (NT), and The Royal Society (JRY). This work used the ARCHER UK National Supercomputing Service (http://www.archer.ac.uk), for which access was obtained via the UKCP consortium and funded by EPSRC grant ref EP/K013564/1. We are also grateful for many valuable discussions with M. Ye and D. Vanderbilt. 
\end{acknowledgments}

\bibliography{library}

\afterpage{\newpage~\newpage}


\pagebreak
\widetext

\includepdf[pages={1}]{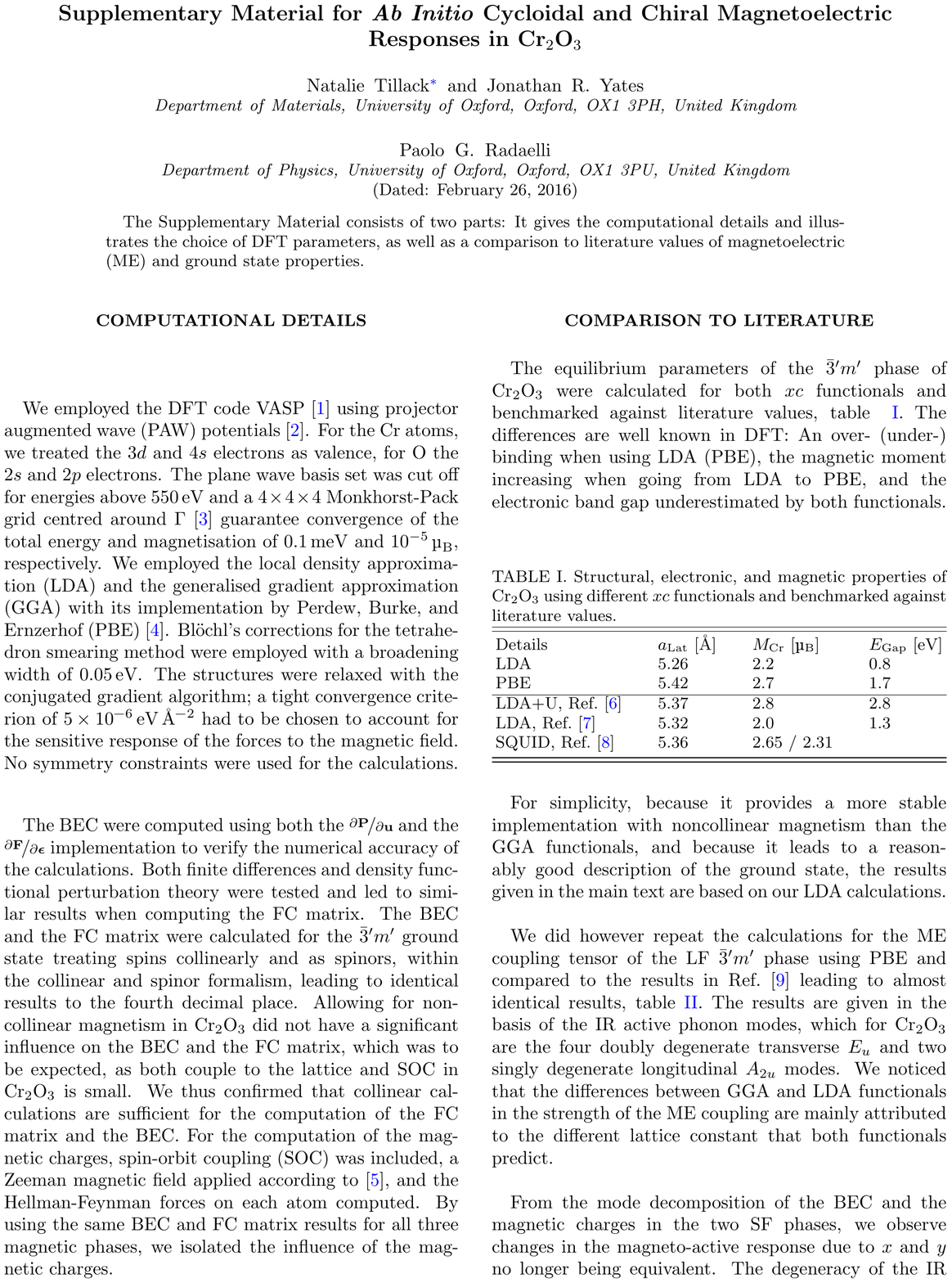}
\afterpage{\newpage}
\includepdf[pages={2}]{si_20160226_NT.pdf}

\end{document}